\documentclass[12pt]{article} 
\usepackage{epsfig,latexsym}   
\usepackage[hang,bf,small]{caption}   
\DeclareGraphicsExtensions{.eps.gz,.eps,.ps,.ps.gz}   
\parskip=\itemsep              
\newcommand{\beq}{\begin{equation}}   
\newcommand{\eeq}{\end{equation}}   
\newcommand{\beqar}[1]{\begin{eqnarray}\label{#1}}   
\newcommand{\eeqar}{\end{eqnarray}}   
\def\npb#1#2#3{    {\it Nucl. Phys. }{\bf B#1} (#2) #3}   
\def\plb#1#2#3{    {\it Phys. Lett. }{\bf B#1} (#2) #3}   
\def\prd#1#2#3{    {\it Phys. Rev. }{\bf D#1} (#2) #3}

\def\zpc#1#2#3{    {\it Z. Phys. }{\bf C#1} (19#2) #3}

\relax   
\begin{document}   
\title{   
{\Large \bf The Dipole Picture and   Saturation in Soft Processes }\\\  
}  
\author{   
{\large ~J.~Bartels,  
\thanks{e-mail:jochen.bartels@desy.de}~~$\mathbf{{}^{a)}}$ 
 ~E.~Gotsman, \thanks{e-mail:   
gotsman@post.tau.ac.il}~~$\mathbf{{}^{b),\,c)}}$ E.~Levin,\thanks{e-mail:   
leving@post.tau.ac.il;\,\,\,levin@mail.desy.de}~~$\mathbf{{}^{b),\,c)}}$   
  M.~Lublinsky,\thanks{e-mail:   
lublinm@mail.desy.de;\,\,\,mal@tx.technion.ac.il}~~$\mathbf{{}^{c)}}$ 
\,and\,U.~Maor \thanks{e-mail:maor@post.tau.ac.il}~~$\mathbf{{}^{b)}}$   
}\\[4.5ex]  
{\it ${}^{a)}$ II Institut f\"{u}r Theoretische Physik,}\\    
{\it Universit\"{a}t  Hamburg, D-22761 Hamburg, GERMANY}\\[2.5ex]  
{\it ${}^{b)}$ HEP Department}\\   
{\it  School of Physics and Astronomy}\\   
{\it Raymond and Beverly Sackler Faculty of Exact Science}\\   
{\it Tel Aviv University, Tel Aviv, 69978, ISRAEL}\\[2.5ex]   
{\it ${}^{c)}$ DESY Theory Group}\\   
{\it 22607, Hamburg, GERMANY}\\[4.5ex]   
}   
   
\date{\today}   
\maketitle   
\thispagestyle{empty}

\begin{abstract}   
 We attempt to describe soft hadron interactions in the framework of  
saturation models, one based upon the Balitsky-Kovchegov non-linear 
equation and another one due to Golec-Biernat and W\"{u}sthoff.
For $pp$, $Kp$, and $\pi p$ scattering the relevant 
hadronic wave functions are formulated, and total, elastic cross-sections, 
and the forward elastic slope are calculated and compared to experimental  
data. The saturation mechanism leads to reasonable reproduction of the  
data for the  quantities analyzed, except for  the forward elastic slope, 
where the predicted increase with energy is too moderate. 
 \end{abstract}   
\thispagestyle{empty}   
\begin{flushright}   
\vspace{-20.5cm}   
TAUP-2729-2002\\   
DESY 02-219\\   
\today   
\end{flushright}   
\newpage   
\setcounter{page}{1}

\section{Introduction}   
\setcounter{equation}{0}   
 
 Understanding the high energy behaviour of total hadronic cross-sections 
within the framework of QCD is one of the intriguing problems of high  
energy 
physics. The main difficulty lies in the fact that presently 
most applications of QCD are based on 
perturbation theory which is only applicable for "hard" processes (i.e. 
it needs a "hard" scale), while hadronic processes near the forward  
direction 
are "soft" and non-perturbative by definition. On the other hand, 
the past few years have seen much activity in the successful application 
of QCD to DIS processes. For values of $Q^2 \; \geq \; 2 \; GeV^2$, 
the use of perturbative QCD seems to be trustworthy. For very small $Q^2$ 
one can rely on Regge theory (e.g.\cite{ZEUS}) which provides a reasonable  
description of the data. The construction of a very promising bridge  
between 
these two theoretical frameworks has been pioneered by the concept of high 
parton densities and saturation. Models based upon this idea have been 
successful in describing 
the DIS cross-section for all values of $Q^2$ and  energies $ x\; \le $  
0.01. 
\cite{GBW,BG,GLLM}. 
 
The goal of this paper is an attempt to apply the dipole picture and the 
physics of high parton densities to soft hadronic cross-sections. 
We want to explore to what extent the high energy hadronic asymptotic 
behaviour can be explained by the saturation hypothesis, which - so far - 
has been tested in the context of deep inelastic scattering at 
small $x$ and in $\gamma-\gamma$ scattering. 
The idea of saturation 
concerns the interactions between partons from different cascades, which  
in the 
linear evolution equations (DGLAP and BFKL) are not included, 
and which become more important with increasing energy. 
The parton saturation phenomenon then introduces  a characteristic  
momentum 
scale $Q_{s}(x)$, which is a measure of the density of the saturated 
gluons. It grows rapidly with energy, and it is proportional to 
 $\frac{1}{x^\lambda}$ \cite{GLR,hdQCD,MV} with $\lambda \simeq 0.2$. 
Parton saturation effects are expected to set in at low values of 
$Q^2$ and $x$, where the parton densities are sufficiently large. 
 
At this stage we do not discern how to relate the dipole picture to the 
additive quark model which has been successful in explaining and 
relating different hadronic total cross sections. 
Instead, we  consider this study as 
being exploratory, and we will not attempt to draw any conclusions 
concerning this rather fundamental issue. 
 
The basis of our endevour is the successful fit \cite{GLLM}  to the 
$F_{2}$ structure 
function data for all values of $Q^2$ and $x \; \le$ 0.01, within the 
framework of 
QCD,  achieved by using an approximate solution to the 
Balitsky-Kovchegov (BK) \cite{BK}  nonlinear evolution equation, and adding a 
correcting 
function to improve the DGLAP behaviour at large $Q^2$. Although  soft 
physics  is not explicitly included, agreement with experiment is found 
for all parameters associated with $F_{2}$, in particular for the 
logarithmic slope 
$\lambda\equiv \partial\ln F_2/\partial(\ln 1/x)$, 
  a value of $\lambda \; \approx \; 0.08$ was obtained 
 at very low $x$ and  $Q^2$ well below $1\, {\rm GeV^2}$, i.e. 
in the saturation region. This agrees with the value of the 
intercept of the  "soft" Pomeron, associated with the Donnachie-Landshoff (DL) 
model \cite{DL}. 
 
In this paper we start from the hypothesis that also in hadron-hadron 
scattering at high energies color dipoles might be the correct degrees of 
freedom, 
even when large transverse distances come into play. 
We start from the well known expression for DIS cross-sections 
\beq \label{1.1} 
 \sigma^{\gamma^{*}p}_{T,L}(x,Q^2) \; = \; \int d^2r_{\perp} dz 
|\psi_{T,L}(Q,r_{\perp},z)|^2 \sigma_{dipole}(x,r_{\perp}) 
\eeq 
where $Q^2$ denotes the virtual photon's four momentum 
squared, $\psi_{T,L}$ its 
wave function, $W^2$ the energy squared in the photon-proton 
system and 
 z, (1-z)  the momentum fraction taken by the quark (antiquark) 
respectively. 
$r_{\perp}$ is the transverse distance between the q and ${\bar q}$, and  
$ x \; = \; \frac{Q^2}{(W^2 \; + \; Q^2)}$. 
There are two main elements in Eq.(\ref{1.1}): (a) the wave function  of 
the virtual photon,  and (b) the 
dipole cross-section, which describes the interaction of the 
$q {\bar q}$ with the proton target, through the exchange of a  gluon 
ladder. In the DIS case the wave function for the virtual photon is well 
known, whereas for the hadron case this is not so. We  discuss 
the question of  the hadronic wave functions in Section 3. 
  
It would be naive for us to expect that our treatment is able to yield 
the complete hadronic cross-section. We have, at least, two reasons for 
this statement: first, at large impact 
parameter we have to include the non-perturbative contributions even for 
the so called "hard" processes \cite{LRREV,KW,MM} since this behaviour is 
defined by the spectrum of hadrons \cite{FROI}; second, 
at present 
the impact parameter dependence of the interaction is only treated 
approximately. In \cite{GBW}, the dipole cross section has a built-in     
sharp cutoff in $b$ at the value of the proton radius; in other 
cases \cite{GLLM} the equation is first solved for b = 0, and then an  
ansatz 
is made regarding factorization and the assumed b dependence of the dipole 
cross section. 
For hadronic interactions the impact parameter dependence is known to be 
important, and neither a sharp (in particular: energy independent cutoff) 
nor the method of calculating  saturation at b = 0, and assuming that the 
$b$-shape does not change with energy, is, at best, a very rough 
approximation to the physical situation. 
 
The content of the paper is as follows. In Section 2 we discuss the 
numerical solution   of the BK 
 equation \cite{BK}, and the changes 
that must be made to adapt this for the calculation of hadron-hadron 
cross-sections. In 
Section 3 we present the details of our calculation. Section 4 is 
devoted to the overall picture, including comparison of the model 
predictions with experimental data. Section 5 contains a discussion of our 
results and our conclusions. In the Appendix we explain why the 
results for the  forward elastic slope are so shallow.


\section{ The Master Equation}   
  \setcounter{equation}{0} 
 
 In \cite{GLLM} an approximate solution to 
the BK non-linear 
evolution equation \cite{BK} was obtained using numerical techniques.  
 Below we  briefly review  
the method used and the main results obtained. For more 
 details of the method of solution  we refer to \cite{GLLM}. 
 
\begin{figure}[htbp]   
\begin{minipage}{9.0cm}   
\epsfig{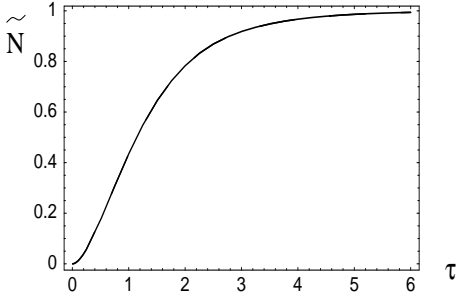}   
\end{minipage}   
\begin{minipage}{7.0 cm}   
\caption{ \it $\tilde N(b=0)$  versus $\tau=r_\perp\, Q_s(x)$. } \label{scal}   
\end{minipage}   
\end{figure}

The solution of the BK equation which we denote by  
$\tilde N$, takes into account the collective phenomena of high parton  
density QCD.  Starting from an initial condition which contains  
 free parameters we have numerically solved the nonlinear evolution    
equation, restricting ourselves to the point $b=0$. The parameters have been  
fitted to the $F_2$ data \cite{GLLM}, and the resulting approximate 
solution is   displayed in Fig. \ref{scal}. 
 
The $b$-dependence of the solution is restored using the ansatz: 
\beq  
\label{Nb} \tilde N(r_\perp,x; b)\,=\,   
(1\,-\,e^{-\kappa(x,r_\perp)\, S(b)})\,,  
\eeq  
where $\kappa$ is related to the $b=0$ solution   
\beq \label{kappa}   
\kappa(x,r_\perp)\,=\,-\,\ln(1\,-\,\tilde N(r_\perp,x,b=0)).  
\eeq   
The Gaussian form for the profile function in impact parameter  
space was assumed i.e. 
\beq\label{sb} 
    S(b_{\perp}) \; = \; \frac{1}{ \pi\; R_{proton}^2} 
exp(- \frac{b^2_{\perp}}{ R_{proton}^2})\,. 
\eeq 
where $R_{proton}^2\;$ =  3.1 $GeV^{-2}$ refers to the radius of the target 
proton. The dipole-proton cross-section (from eq.(\ref{DN})   
of \cite{GLLM}) is given by: 
\beq 
 \label{DN} 
\sigma_{dipole}(r_\perp,x) \; = \; 2 \; \int \; d^2b_\perp\, 
 \tilde N( r_{\perp}, b_{\perp}, x)\,.  
\eeq  
  
Another popular  saturation model was proposed by Golec-Biernat and  
W\"{u}stoff  \cite{GBW} which we will denote by GBW.
The following dipole cross-section is assumed to be: 
\beq\label{dGBW} 
 \hat{\sigma}(r_{\perp},x)_{dipole} \; = \; \sigma_{0}[1 \; -  \;  
exp(-\frac{r_{\perp}^2}{4 R^{2}_{0}})] 
\eeq 
with $R^{2}_{0}(x)[GeV^{-2}] \; = \; (\frac{x}{x_{0}})^{\lambda}$. 
The values of the parameters which were determined by fitting to DIS data  
at  
HERA for $x \le 0.01$, are: $\sigma_{0} \; = $ 23 mb, $ \lambda \; =  
\;0.29$ 
and $x_{0} \; = \; 3\times10^{-4}$.  
 The $r_{\perp}$ dependence is taken as Gaussian, which leads 
to a constant cross-section $\sigma_{0}$ for large $r_{\perp}$ (or small 
$Q^2$) i.e. for  "soft" interactions. 
 
The  dipole cross-section of the  GBW saturation model and $\tilde N$ 
are closely related. Both models include the effects of gluon saturation, 
preserve unitarity, and describe the physics associated with  
"long distances". Whereas $\tilde N$ has some support from QCD, $\hat{\sigma}_{dipole}$ of the GBW model has more the character of a phenomenological model.

Unlike the GBW $\hat{\sigma}_{dipole}$, the dipole cross section  
obtained from the  
solution of the BK equation  is not saturated as a function of $x$. 
 This emanates from   the  integration over $b$.  
With the assumed Gaussian profile function it  leads to a logarithmic  
growth with decreasing $x$.
  The Froissart-like behaviour $\sigma_{dipole} 
\,\propto\,\ln^2(1/x)$ \cite{FROI} is crucially dependent on the fact that   
   the large impact  
parameter behaviour of the dipole cross-section 
 is exponential, 
rather than Gaussian.

Starting from our dipole cross sections we obtain 
  our master equation for the hadron-proton cross-section 
\beq \label{sighp} 
 \sigma_{H-proton}(x) \; = \; \int d^2 r_\perp   
 | \psi_{H}(r_\perp) |^2\,\sigma_{dipole}(r_\perp,x)\,, \eeq 
where $\psi_{H}(r_\perp)$ represents the wave function of the  
hadron which  
scatters off the target proton. The form taken for 
 $\psi_{H}(r_\perp)$ is  
discussed in the next section.  
 
 For both saturation models ((\ref{DN}) and (\ref{dGBW})) the energy  
dependence of the hadron-proton cross section  
enters only through $x$-dependence of the dipole cross section, 
the latter being adjusted or constructed to describe DIS data of the $F_2$  
structure function. The way in which the the $x$-dependence of the dipole
 cross section determines the energy dependence of the hadron cross sections
is strongly influenced by the $b$-slope.
Note, however, that there is no $x$ in hadron-hadron collisions. In order to  
relate $x$ to the energy of the process we will need to introduce an additional 
nonperturbative scale, denoted below as $Q_0^2$.

\section{Details of Calculation} 
\setcounter{equation}{0} 
 
\subsection{Hadronic wave function} 
 There is no established method  for calculating hadronic wave functions  
within the framework of QCD. The Heidelberg group Dosch et al.\cite{Dos}   
using  
the stochastic vacuum model, have calculated hadronic cross-sections,  
after making an ansatz regarding the form of the hadronic wave function 
 $\psi_{H}(r )$. 
We adopt their ansatz and utilize hadronic wave functions of  similar  
shape to those 
 used in \cite{Dos}.  
 
 Based on experimental evidence of the  
flavour dependence of hadronic cross-sections, which decrease with  
increasing number of strange quarks,   the authors of \cite{Dos} 
 hypothesized that the cross sections  
depend on the sizes of the hadrons  in the process.  
 
   For the hadron transverse wave function we take 
 a simple Gaussian 
form, the square of the wave function is given by 
 \beq \label{3.1} 
 |\psi_{M}(r_{\perp})|^2 \; = \;  \frac{1}{\pi S_{M}^2} 
exp(- \frac{r^{2}_{\perp}}{S_{M}^2}) 
 \eeq 
where $S_{M}$ is a parameter related to the meson size. We have used 
$S_{\pi} \; = \; 1.08$ fm and 
$S_{K} \; = \; 0.95 \; $   fm.  These  $S_M$ were found 
from experimental values for the electromagnetic radii, namely, $R_{\pi}= 
0.66\pm 0.01$ fm and $ R_{K} = 0.58 \pm 0.04 $ fm \cite{RAD}. For  meson 
wavefunction of the form (\ref{3.1}) $S_M=\sqrt{\frac{8}{3}} R_M$. 
 
The proton's wave function squared  is given by 
\beq \label{3.2} 
 |\psi_{p}(r_{1 \perp},r_{2 \perp})|^2 \; = \; \frac{1}{(\pi S_{p})^2} 
exp(- \frac{r_{1 \perp}^2 \; + \; r_{2 \perp}^2}{S_{p}^2}) 
\eeq 
where $S_{p} \; = \; 1.05 $ fm,  which corresponds to $R_p = 0.862 \pm 
0.012 $ fm \cite{RAD}.  For proton 
wavefunction of the form (\ref{3.2}) $S_p=\sqrt{\frac{3}{2}} R_p$.

  For meson-proton scattering, the meson is treated as a 
quark-antiquark pair (i.e. a colour dipole), and therefore the calculation 
follows that of  DIS, i.e. the interaction of a colour dipole with a 
proton target, with the meson wave function replacing that of the virtual 
photon. However, for the scattering of a baryon projectile, we  represent 
the baryon as constituted of two colour dipoles, one dipole formed around 
two quarks, and the second dipole from the center of mass of these two 
quarks to the third quark in the baryon.  Generally speaking the parameter 
$S_p$ in (\ref{3.2}) can be different for these two dipoles. For example, in 
the non relativistic additive quark model (AQM) we expect $S_p$ for the first  
dipole to be larger by the factor 4/3 compared to the second dipole. In  
(\ref{3.2}), for simplicity, they are choosen to be identical.  
It will be shown below  
that the hadron cross sections are mostly determined by the saturation domain 
where the sensitivity to the wave function and, in particular,  
to the choice of $S_p$ is quite weak.
 
In AQM, the $\frac{2}{3}$ ratio between  $\pi$-p and p-p 
cross-sections  is due to quark counting.  In our model, at very high  
energies when the dipole cross section is independent of the dipole size, 
the predicted  
 ratio is $\frac{1}{2}$, given by the number of dipoles in the 
pion relative to those in the proton.  
 In our approach the high 
energy interaction is blind with respect to the flavours of the interacting 
quarks, and the ratio 
$\sigma^{Mp}_{tot}/\sigma^{pp}_{tot} = 1/2$, 
seems to disagree with the data. 
 Experimentally, at an energy of 
$\sqrt{s}$ = 20 GeV, 
$\sigma_{tot}^{\pi p} / \sigma_{tot}^{p p} \approx $ 0.6. 
 We expect  the secondary Regge trajectories 
 to give  a smaller contribution to $K^+p$ interaction, 
 as there are no resonances in  
the 
$s$ channel of this reaction, the same is also true 
   for proton-proton scattering.  
The predicted ratio  $\sigma^{K^+p}_{tot}/\sigma^{pp}_{tot}= 1/2$ is in 
reasonably good agreement with the experiment data. 
 
\subsection{Method of  Calculation} 
 
   All the parameters in $N(r_\perp,b_\perp, x)$ were taken from the fit 
of 
 \beq\label{F2} 
 F_{2}(Q^2,x) \; = \; \frac{Q^2}{4 \pi^2 \alpha_{elem}}\, 
 \sigma^{\gamma^{*}p}(Q^2,x)  
\eeq 
 
  made to the experimental DIS data, 
(see \cite{GLLM} for details). In the DIS case the variable $x$ is well  
defined 
in terms of $Q^2$ and $W^2$, in the hadronic case we  redefine 
$x$ to be  $ x \; = \; \frac{Q^2_0}{s}$, where s denotes the energy  
squared  
in the center of mass system of the hadrons, and $Q^2_0$ is a parameter 
which we  adjust to be compatible with the data. 
 The value of $Q^2_0$ is determined by the longitudinal part of 
the wave function, for which at present we do not 
 have a reasonable model. In general we would expect the scale $Q_0$ to  
increase with increasing hadron masses, and thus vary from hadron to hadron. 
In this study we will fit this parameter separately for each projectile  
hadron.
 
In the colour dipole picture (which is equivalent to two gluon exchange), 
one can hopefully only reproduce the asymptotic energy dependence i.e. the  
 Pomeron contribution. At lower energies where most of the data 
for meson baryon scattering is available, there are also  contributions  
from secondary trajectories. The evaluation of these contributions is 
beyond  the scope of our 
model.

\section{Comparison of Our Model Predictions with Data} 
\setcounter{equation}{0} 
 
\begin{figure}[htbp] 
\begin{tabular}{c c} 
(a) & (b) \\ 
 \epsfig{file=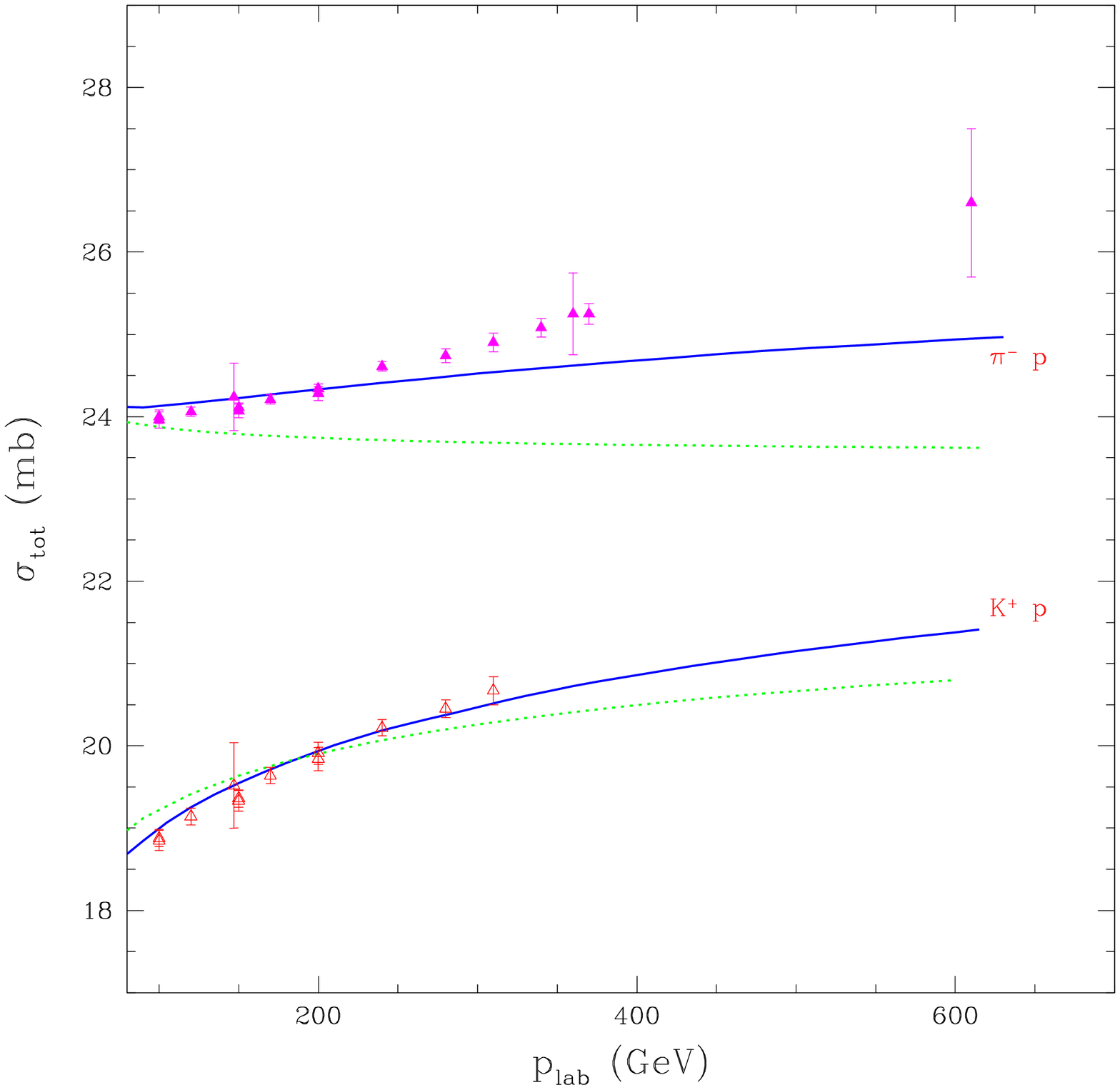,width=85mm, height=95mm}& 
 \epsfig{file=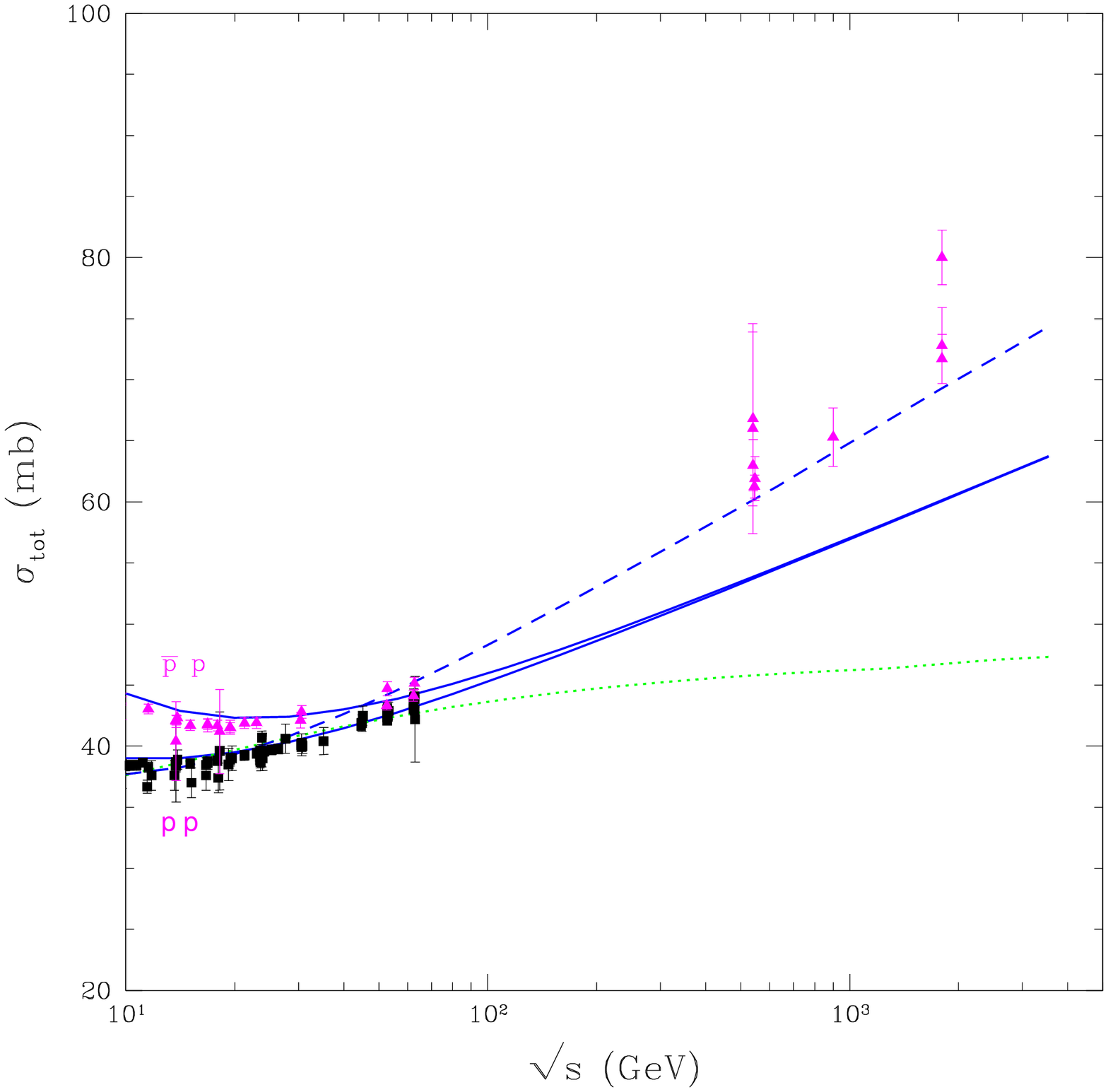,width=85mm, height=95mm} 
\end{tabular} 
  \caption[]{ (a) \it $K^{+}p$ and $ \pi^{-} p$ total cross-sections. 
The full line is the prediction of our model and the dotted line using the 
Golec-Biernat W\"{u}stoff dipole.\\ 
(b)  ${\bar p}p$ and $ p p$ total cross-sections. The full 
lines are the model predictions with a Gaussian profile. The dashed line 
is the result of using a $K_{1}$ profile, and the dotted line using the 
Golec-Biernat W\"{u}sthoff dipole.\\ 
Data compilation from Ref. \cite{DATA} } 
    \label{KpXS} 
\end{figure}

 Our model contains a parameter $Q_{0}^2$, (which can be considered as a  
scale factor) from the definition of  
$x \; = \; \frac{Q^2_{0}}{W^2}$, in analogy with the variable in DIS, 
$\;  x \; \approx \; \frac{Q^2}{W^2}$. The energy dependence of the  
hadronic  
cross-sections can be adjusted by choosing a suitable value for this  
variable. We found that the value $Q_{0}^2 \; = \;3.5 \times 10^{-3}\;  
GeV^2$, 
 gives 
  good agreement with the 
data for $\sigma_{tot}(K^+p)$. See Fig. \ref{KpXS} (a). The $K^{+}p$ 
channel was chosen as it is exotic, having 
 no resonances in s channel and therefore by  
duality the secondary Regge  trajectory contributions are small.   
By replacing the wave function of the $K^{+}$ meson  in Eq.(\ref{3.1}) 
by that of the $\pi$ meson, we obtain $\sigma_{tot}$ for the $\pi^{-}p$. 
We display our prediction and the data in Fig. \ref{KpXS} (a).  
We have chosen to 
show $\sigma_{tot}(\pi^{-}p)$ compared to data, as there are more data 
in this channel than in the $\pi^{+}p$ channel. The cross-sections 
 $\sigma_{\pi^{+}p}$ and $\sigma_{\pi^{-}p}$ only 
differ  due  to the contribution of the secondary trajectories. 
The parameters used for $\sigma_{tot}(\pi^{-}p)$ 
 are  $Q_{0}^2 \; = \; 1 \times 10^{-3}\; GeV^2$, 
and a Regge contribution of $ 27\times(\frac{s}{s_{0}})^{-0.45}$ mb 
($s_{0} \; = \;  1\; GeV^2$) which has a smaller residue than that suggested  
by the DL model.

We note that the predicted energy dependence for  
 $\sigma_{tot}(\pi^{-}p)$ is more moderate than the data, and at 
an energy of $p_{lab}$ = 400 GeV, we underestimate the experimental 
data by approximately 7 \%.  
 
We compare our results with those of the GBW model, by 
replacing $\sigma_{dipole}$ in Eq. (\ref{sighp}), by the GBW 
dipole ${\hat \sigma}(x,r_{\perp})_{dipole}$ given in Eq. (\ref{dGBW}). 
We wish to stress that the GBW saturation model  
 was formulated for, and applied to DIS reactions \cite{GBW}. 
The responsibility of extending the model to hadronic interactions is  
ours. 
The GBW  model does not contain any explicit $b$-dependence, but one 
can consider the constant cross-section to be  
the result after integration over the impact parameter (with a sharp impact  
parameter cutoff put into the exponent in (2.5)). It is clear that at 
asymptotic energies (where one can neglect the contribution of secondary 
trajectories), the GBW cross-section for $\sigma_{tot}(Mp) \; \le$ 23 mb,   
and $\sigma_{tot}(Bp) \; \le$ 46 mb, as we have one (two) dipoles 
interacting 
with the proton target for Mp (Bp) scattering.  
In Fig. \ref{KpXS} (a) we show the  
result 
(dotted line) 
obtained using the GBW form for the dipole Eq. (\ref{dGBW}) for the $K^{+}p$ 
channel, with $Q^{2 \; (GBW)}_{0}$ = 0.27  $GeV^2$, and using the same  
form of the hadronic wave function as discussed in Section (3.1).  
 For $\pi^{-}p$ channel we used the almost maximal possible dipole  
cross section, put $Q^2_0=5\times 10^{-4}$ $GeV^2$, and add a secondary  
trajectory with $11 \times(\frac{s}{s_{0}})^{-0.45}$ mb.
Due to the arguement  presented above, 
   the GBW model cannot be adjusted 
to these data at all.  
 
For  $\sigma_{tot}({\bar{p}p})$ (see Fig. \ref{KpXS} (b)) 
 the  energy dependence predicted by the model 
 is not as steep as the experimental data, yielding a value for 
$\sigma_{total}({\bar  p}p)\; \approx \; $ 65 mb (instead of 72 mb) at 
 Tevatron energies. i.e. a deficit of 10 \%, however, 
this is  over a much wider 
energy range than in the $\pi^{-}p$ case. For $\sigma_{total}(p p)$ 
(where data is only available over a  narrower range of energy) 
we achieve a very good reproduction of the experimental data, this is  
displayed in Fig. \ref{KpXS} (b). 
For the pp and $p {\bar p}$ channels  we take $Q^2_{0}$ = 0.03 $GeV^2$, 
and following \cite{DL} have a Regge contribution of 
 $ 98.4\times(\frac{s}{s_{0}})^{-0.45}$ mb for 
$\sigma_{tot}( {\bar p}p)$ and 
 $ 56\times(\frac{s}{s_{0}})^{-0.45}$ mb for $\sigma_{tot}(p p)$.

  To get a handle on the theoretical uncertainties of our 
treatment we have also calculated $\sigma_{tot}({\bar p} p)$, using 
a profile function 
\beq 
S(b) \; = \; \frac{2}{\pi R_{proton}^2} 
(\frac{\sqrt{8}b}{R_{proton}}) K_{1} 
(\frac{\sqrt{8} b}{R_{proton}}) 
\eeq 
which corresponds to the Fourier transform of the "dipole" form factor 
in the momentum transfer representation: 
 \beq 
F_{"dipole"}(t) \; = \; \frac{1}{(1 \; -\; \frac{R_{proton}^2 \;t}{8})^2}\,. 
\eeq 
The result of our calculations with the "dipole" form factor   
  for $\sigma_{tot}( {\bar p}p)$ is shown  by the dashed  
 line in Fig. \ref{KpXS} (b). For this calculation we took 
 $Q^2_{0}$ = 0.06 $GeV^2$. 
For the comparison with the $F_{2}$ data \cite{GLLM}, a value of 
 $R^2_{proton} \; = \; 4.46 \; GeV^{-2}$ was used.

We repeat the same procedure  for the $ {\bar p} p$ channel 
for the GBW model as we did for $\sigma_{tot}(K^+ p)$ explained 
above, now 
 with  
parameters 
$Q^{2 \; (GBW)}_{0}$ = 0.7  $GeV^2$, and a Regge contribution of 
 $ 20\times(\frac{s}{s_{0}})^{-0.45}$ mb, (adjusted to the data), and 
 taking the same form  
for the baryon 
wave function. The results are shown as a dotted line in Fig. \ref{KpXS} (b).

\begin{figure}[htbp] 
 \epsfig{file=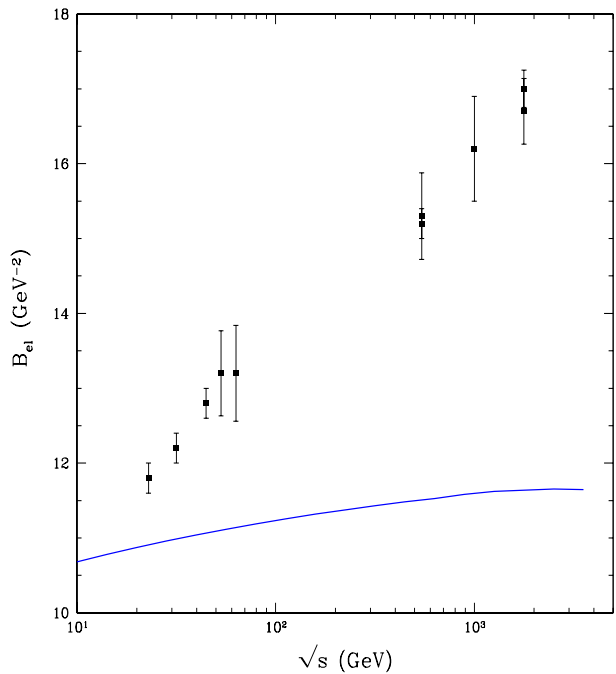,width=85mm, height=95mm} 
  \caption[]{\it ${\bar p}p$  forward elastic slope. 
Data compilation from Ref. \cite{DATA} } 
    \label{bslop} 
\end{figure}

 We  also calculate the forward slope of the elastic cross-  
section i.e. B, which is defined as 
$$ \frac{d \sigma}{dt} \; = \; \frac{d \sigma}{dt}_{|_{t=0}}\,\, e^{-Bt}$$ 
and is related to the sizes of the particles participating in the  
reaction.  $B \; = \; B_{0} \; + \; B^{'}$ where 
\beq\label{slop} 
B^{'}\; = \; \frac{ \int d^2 r_\perp   
 | \psi_{H}(r_\perp) |^2\,\, b^2_{\perp}\,\; N(r_\perp,b_\perp,x) \, 
\; db_{\perp}^2} 
{\sigma_{tot}} \; = \;\frac{1}{2}\, < \; b_{\perp}^2 \; >\,. 
\eeq 
In Fig. \ref{bslop} we display 
  $B \; = \; B_{0} \; + \; B^{'}$   
with $B_{0}$ = 7.8 $GeV^{-2}$. $B_{0}$ is related to the formfactors of the  
hadrons. Its value was chosen with an eye 
on the data, and is close to that used by Schuler and  
Sj\"{o}strand \cite{SS}.

The results we obtain are  
disappointing, but understandable and demonstrate the weak point of our  
model viz. the assumption of the oversimplified form for the impact parameter 
dependence of the amplitude (see Appendix).   
The  elastic slope (unlike $\sigma_{tot}$) 
is sensitive to the $b_{\perp}$ distribution, and our assumption that the  
major contribution comes from small values of $b_{\perp}$ is obviously 
wrong. We will expand on this difficulty in Section 5 and the Appendix.

\begin{figure}[htbp] 
\begin{tabular}{c c} 
(a) & (b) \\ 
\epsfig{file=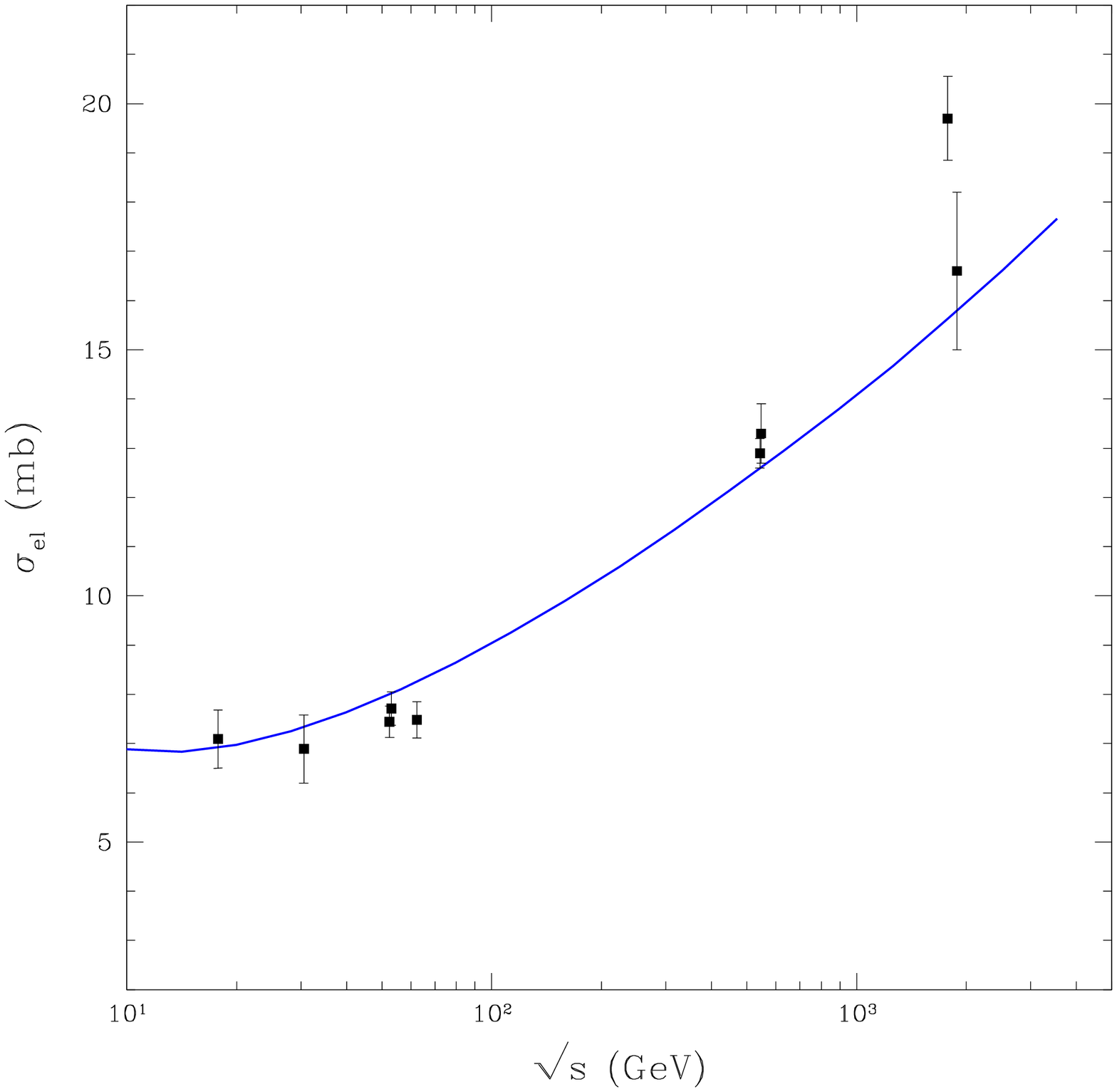,width=85mm, height=95mm}& 
 \epsfig{file=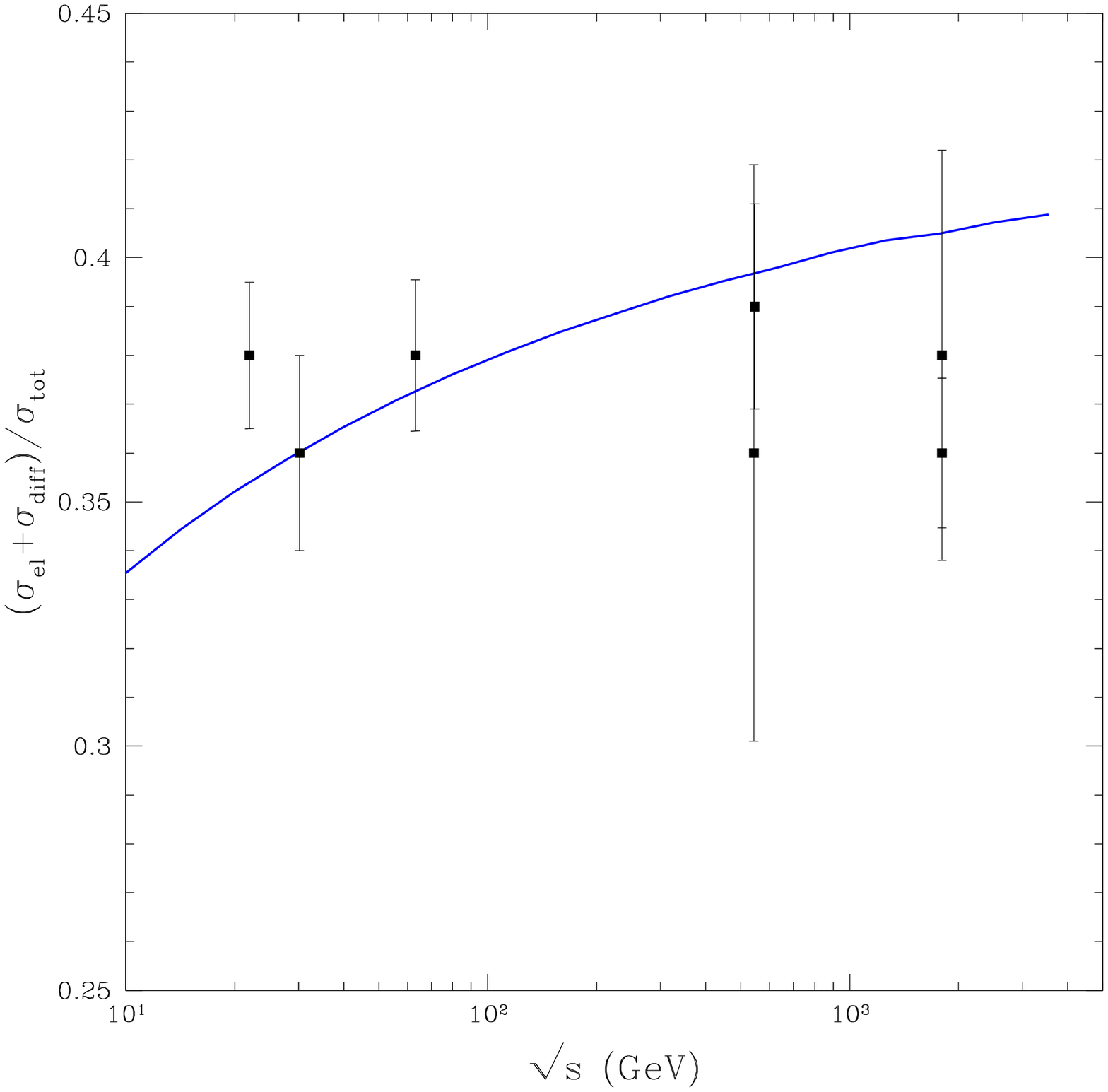,width=85mm, height=95mm}  \\ 
\end{tabular} 
  \caption[]{\it For ${\bar p}p$ scattering; 
(a) elastic cross-section 
(b) the ratio  $ R_{D} $. \\ 
Data compilation from Ref. \cite{DATA}, \cite{GOL} } 
    \label{elrat} 
\end{figure}

Using the relation 
$$ 
\sigma_{elastic} \; = \; \frac{(\sigma_{tot})^2}{16 \; \pi \; B} $$ 
we check our model's predictions for $\sigma_{elastic}$ 
for the ${\bar p}p$ channel, where the data extends to high energies. 
The model  produces a very good  
description of the elastic ${\bar p}p$ cross-section 
 (see Fig. \ref{elrat}(a)). In $\sigma_{elastic}$, the deficiency in the energy 
rise of $\sigma_{tot}$ is to some extend compensated by the inadequacy of $B$. 
  
 We  find it of interest to investigate how close we are to a black 
disc picture for dipole-proton scattering, which we  do by utilizing  
the 
  Pumplin bound \cite{PUM}, which follows from Unitarity considerations 
and can be written as: 
$$ R_{D} \; = \;  \frac{\sigma_{elastic} \; + \; \sigma_{diff}}{\sigma_{tot}} 
\; = \; \frac{ \int d^2 r_\perp d^2 b_\perp 
 | \psi_{H}(r_\perp) |^2 
  N^2({r_{\perp}, b_{\perp}}, x)}{\sigma_{tot}} 
 \le \; \frac{1}{2}\,. $$ 
In Fig. \ref{elrat}(b) we display the ratio $R_{D}$ and the 
experimental data.  The model's predictions agrees with the data, and  
suggests that even at Tevatron energies we are 20\% away from the black 
disc limit of $\frac{1}{2}$.

\section{Discussion and Conclusions}   
\setcounter{equation}{0} 
 
Our treatment has two parameters: \\ 
(a) $R_{h}^{2}$ which is taken from the 
electromagnetic radius of the hadron (following the Heidelberg 
prescription);\\ 
(b) the parameter $Q_{0}^{2}$, (introduced after Eq. (\ref{F2}), 
  is adjusted by comparing with the 
data for the different channels.  It is worth mentioning that,  
as expected, the obtained values for $Q_0$ reflect the mass hierarchy   
of the projectile hadrons.
\\ 
In addition, we require the contribution of secondary trajectories at  
lower energies. 
 
We would like to emphasis that our approximate solution to the BK  
equation 
\cite{GLLM} is obtained for impact parameter b = 0, and then an 
ansatz is made regarding the b dependence of the profile function S(b). 
In the original fit to the $F_{2}$ data \cite{GLLM}, S(b) was 
 taken to be a Gaussian, 
(which is equivalent to assuming that the dependence upon t, the 
momentum transfer squared, is exponential). 
It is this Gaussian shape of the profile function which produces a 
cross-section with a 
$\ln(\frac{1}{x})$ dependence. We have shown that taking a 
 "dipole" behaviour in the $t$ representation we would have a 
profile function $ S(b) \; \sim  \; \frac{b}{R}K_{1}(\frac{b}{R})$, 
and the resulting cross-sections would asymptotically have a 
$\ln^{2}(\frac{1}{x})$ dependence. A  reflection of this fact 
is that the energy dependence  
 of the cross section (see dashed line of Fig. \ref{KpXS}) 
is much steeper,  which is more in accord with the data. 
 Since the original fit to the DIS 
$F_{2}$ data was made with a Gaussian profile in b space, for consistency 
one should redo the fit with a different profile. This is a task for the 
future. We summarize our main results: 
\begin{itemize} 
\item We obtain a reasonable value of the total cross-sections. 
\item The predicted energy increase is too moderate. Both the value of the 
      cross section and the energy dependence depend on the assumed 
      $b$-dependence of the profile function. 
      For illustration we 
      have compared with the saturation model of \cite{GBW}: 
      because of the sharp cutoff in $b$ built into this model, the energy 
      dependence is even weaker. This emphasizes the need to improve, 
      in all saturation models the $b$-shape of the profile function. 
\item The difference between $\sigma_{K^+ p}$ and $\sigma_{ \pi p}$ 
      at high energy needs an explanation, which cannot be answered within 
      the framework of our model. 
\item The slow increase of the slope B, is the result of having an almost 
      black disc picture for  dipole - proton scattering \cite{3ch} and not 
      a consequence of a particular form for the $b$-profile. 
\item $ R_{D} \; = \; \frac{ \sigma_{elastic} \; + \;  
\sigma_{diff}}{\sigma_{tot}}$ 
      tends to $\frac{1}{2}$, which is the black disc limit.  
      We reproduce the experimental values for this ratio, 
      which we consider as a success of the model. 
\end{itemize}

In general the approach works better than one would expect, even in the 
region of long distances.  The results obtained are not very sensitive 
to the input parameters of the projectile hadron, namely the wavefunction  
and the parameter $Q_0^2$. Our use of the saturation models has   
some predictive power, provided we have enough information about 
projectile wave function.
 
Nevertheless, there are serious shortcomings 
in applying the dipole picture with the concept of saturation 
to hadronic cross sections. The crucial feature seems to be the 
$b$-dependence of the dipole cross section which needs further  
investigation. 
On a more fundamental level, it is not clear 
at all, how to relate the dipole picture to the additive quark model: 
at present these two approaches look almost orthogonal to each other.

\section{Acknowledgments} 
 
 We would like to thank Claude Bourrely and Jacques Soffer 
 for their help in  sending us 
their  files of the cross-section data.  We thank Eran Naftali for useful 
remarks.   
Two  of us (E.G. and E.L.) thank the DESY Theory Division for their 
hospitality. 
This research was supported in part by the BSF grant \# 9800276, and by 
the 
GIF grant \# I-620-22.14/1999, and by the Israel Science Foundation, 
founded by the Israeli Academy of Science and Humanities.

\section{Appendix} 
\setcounter{equation}{0}

 Our results can be easily checked in the asymptotic limit where we  
have (for simplicity of argument the following discussion is presented at  
some fixed $r_\perp $, while $r_\perp $ integrations are implicitly assumed): 
\begin{equation} 
\sigma_{Mp} \; = \; 2 \pi \; \int^{b^2_{0}(x)} \; db_{\perp}^2 \; 
= \; 2 \pi b^{2}_{0}(x) 
\end{equation} 
where $b^2_{0}(x)$ is a result of our ansatz on the $b_{\perp}$ behaviour 
viz. 
\begin{equation} 
\sigma_{dipole -  p}(r_{\perp}, x; b_{\perp}) \; = \; 
2\,(1 \; - \; e^{-\Omega/2}) 
\end{equation} 
 
with 
$$\Omega \; = \; \, \ln \; [1 \; \; -  
\; N(r_{\perp},x; b_{\perp}=0)]\, 
e^{-\frac{b_{\perp}^2}{R^2}}\,. $$ 
$b_{0}^2(x)$ can be calculated from the equation 
$$ \frac{\Omega(r_{\perp}, x; b_{\perp} = b_{0}(x))}{2} \; = \; 1 $$ 
i.e. 
$$b^2_{0}(x) \; = \; R^2\, \ln \; [\frac{1}{2}\, 
\ln(1 \; - \;  N(r_{\perp},x; b_{\perp}=0)] $$ 
where 
$\sigma_{\pi p}^{tot} \; = \; 2 \pi b^2_{0}(x)$ 
and $\sigma_{pp}^{tot} \; = \; 4 \pi b^2_{0}(x)$. The factor 2 between 
the cross-sections is due to the fact that we have 
 two dipoles in the proton and one in the meson. 
 
The elastic slope is given by Eq. (\ref{slop}) 
\beq 
B^{'}\, = \; 
\frac{b^4_{0}(x)/2}{2 \,b^2_{0}(x)} \; = \; \frac{b^2_{0}(x)}{4}\,.  
\eeq   
The energy dependence of $B^{'}$ can now be calculated 
\beq 
\frac{d B^{'}}{d \; \ln(1/x)} \; = \; \frac{1}{4} \frac{d b^2_{0}(x)} 
{d \; \ln(1/x)} \; = \; \frac{1}{16 \pi} \frac{d \sigma_{pp}^{tot}(x)} 
{d \; \ln(1/x)}\,. 
\eeq       
>From this formula it is clear that we have 
  an increase of B with $\frac{1}{x}$ which is 
slower than that of $\sigma^{tot}_{pp}(x)$, and obviously 
not in accord with  the experimental data. Though the above argument was  
presented for the Gaussian profile, similar conclusions would be obtained 
for an alternative profile function.

\end{document}